\documentclass[aps, prl, twocolumn, preprintnumbers, superscriptaddress, showpacs, amssymb]{revtex4-1}

\usepackage{graphicx}% Include figure files
\usepackage{dcolumn}% Align table columns on decimal point
\usepackage{bm}% bold math
\begin{document}

% Use the \preprint command to place your local institutional report
% number in the upper righthand corner of the title page in preprint mode.
% Multiple \preprint commands are allowed.
% Use the 'preprintnumbers' class option to override journal defaults
% to display numbers if necessary
\preprint{Ver. 7sub}¡¡

%Title of paper
\title{Accurate determination of the Fermi surface of tetragonal FeS via quantum oscillation measurements and quasiparticle self-consistent \textit{GW} calculations}

% repeat the \author .. \affiliation  etc. as needed
% \email, \thanks, \homepage, \altaffiliation all apply to the current
% author. Explanatory text should go in the []'s, actual e-mail
% address or url should go in the {}'s for \email and \homepage.
% Please use the appropriate macro foreach each type of information

% \affiliation command applies to all authors since the last
% \affiliation command. The \affiliation command should follow the
% other information
% \affiliation can be followed by \email, \homepage, \thanks as well.
\author{Taichi Terashima}
\affiliation{National Institute for Materials Science, Tsukuba, Ibaraki 305-0003, Japan}
\affiliation{Department of Physics, Tohoku University, Sendai 980-8578, Japan}
%\email{TERASHIMA.Taichi@nims.go.jp}
\author{Naoki Kikugawa}
\affiliation{National Institute for Materials Science, Tsukuba, Ibaraki 305-0003, Japan}
\author{David Graf}
\affiliation{National High Magnetic Field Laboratory, Florida State University, Tallahassee, FL 32310, USA}
\author{Hishiro T. Hirose}
\author{Shinya Uji}
\affiliation{National Institute for Materials Science, Tsukuba, Ibaraki 305-0003, Japan}
\author{Yoshitaka Matsushita}
\affiliation{National Institute for Materials Science, Tsukuba, Ibaraki 305-0047, Japan}
\author{Hai Lin}
\author{Xiyu Zhu}
\author{Hai-Hu Wen}
\affiliation{Center for Superconducting Physics and Materials, National Laboratory of Solid State Microstructures and Department of Physics, National Center of Microstructures and Quantum Manipulation, Nanjing University, Nanjing 210093, China}
\author{Takuya Nomoto}
\affiliation{Department of Applied Physics, University of Tokyo, Bunkyo, Tokyo
113-8656, Japan}
\author{Katsuhiro Suzuki}
\affiliation{Research Organization of Science and Technology, Ritsumeikan University, Kusatsu, Shiga 525-8577, Japan}
\author{Hiroaki Ikeda}
%\email{hikeda.uji@gmail.com}
\affiliation{Department of Physics, Ritsumeikan University, Kusatsu, Shiga 525-8577, Japan}

%Collaboration name if desired (requires use of superscriptaddress
%option in \documentclass). \noaffiliation is required (may also be
%used with the \author command).
%\collaboration can be followed by \email, \homepage, \thanks as well.
%\collaboration{}
%\noaffiliation

\date{\today}
\begin{abstract}
We perform de Haas-van Alphen measurements and quasiparticle self-consistent \textit{GW} (QS\textit{GW}) calculations on FeS.
The calculated Fermi surface (FS) consists of two hole and two electron cylinders.
We observe all the eight predicted FS cross sections experimentally.
With momentum-independent band-energy adjustments of less than 0.1 eV, the maximum deviation between the calculated and observed cross sections is less than 0.2\% of the Brillouin zone area for $B \parallel c$.
%An essential ingredient of this success is that the three dimensionality of the electronic structure is reduced in the QS\textit{GW} calculations compared to calculations density-functional-theory ones.
The carrier density is $\sim$0.5 carriers/Fe.
The mass enhancements are nearly uniform across the FS cylinders and moderate, $\sim$2.
The absence of a third hole cylinder with $d_{xy}$ character is favorable for the formation of a nodal superconducting gap.
%The present results show that the QS\textit{GW} method is a powerful tool to study electronic structures of semimetallic iron pnictides/arsenides and provide a frame of reference for comparison between FeS and exotic FeSe.
\end{abstract}

% insert suggested PACS numbers in braces on next line
%\pacs{74.70.Xa, 74.25.Dw, 74.25.Jb, 71.18.+y}
% insert suggested keywords - APS authors don't need to do this
%\keywords{}

%\maketitle must follow title, authors, abstract, \pacs, and \keywords
\maketitle

% body of paper here - Use proper section commands
% References should be done using the \cite, \ref, and \label commands
%\section{}
% Put \label in argument of \section for cross-referencing
%\section{\label{}}
%\subsection{}
%\subsubsection{}

% If in two-column mode, this environment will change to single-column
% format so that long equations can be displayed. Use
% sparingly.
%\begin{widetext}
% put long equation here
%\end{widetext}

\newcommand{\ud}{\mathrm{d}}
\def\degree{\kern-.2em\r{}\kern-.3em}

\section{intoroduction}
Tetragonal FeS has the same PbO-type structure as FeSe.
Lai \textit{et al}. succeeded to synthesize high-quality tetragonal FeS with a superconducting transition temperature $T_c$ = 5 K using a hydrothermal method \cite{Lai15JACS}, which opened up an opportunity to compare FeS and FeSe.

FeSe is perhaps the most mysterious iron-based superconductor.
Unlike iron-arsenide parent compounds LaFeAsO \cite{Kamihara08JACS} and BaFe$_2$As$_2$ \cite{Rotter08PRL, Sasmal08PRL}, FeSe does not order magnetically at ambient pressure \cite{McQueen09PRL}, although an antiferromagnetic order can be induced by a moderate pressure of $\sim$10 kbar \cite{Bendele10PRL, Bendele12PRB, Terashima15JPSJ}.
It exhibits only a nematic transition, i.e., a tetragonal-to-orthorhombic transition at ambient pressure and becomes superconducting below $T_c$ = 8 K \cite{Hsu08PNAS}.
The onset of superconductivity can be enhanced up to $\sim$37 K by high pressure \cite{Mizuguchi08APL, Medvedev09Nmat} and, moreover, above 50 K in the form of single-layer films \cite{Wang12CPL}.
The nematic order has a profound influence on the electronic structure, as evidenced by the anomalously small Fermi surface \cite{Maletz14PRB, Terashima14PRB, Audouard15EPL, Watson15PRB, Watson15PRL}: the carrier density is only 0.01 carriers/Fe, more than one order-of-magnitude smaller than predicted by band-structure calculations \cite{Terashima14PRB}.

On the other hand, FeS does not show nematicity: it remains tetragonal down to zero temperature \cite{Lai15JACS, Pachmayr16ChemCommun}.
Band-structure calculations predict a quasi-two-dimensional (Q2D) electronic structure similar to those in iron-pnictide superconductors or their parent compounds in the paramagnetic phase \cite{Yang16PRB, Terashima16PRB_FeS}.
FeS is thus a perfect reference compound in studying anomalous properties of FeSe.
It is also to be noted that there is a debate about whether the superconducting gap in FeS is a full gap \cite{Holenstein16PRB, Kirschner16PRB} or a nodal one \cite{Xing16PRB, Ying16PRB, YangXiong16PRB}.
%$\mu$SR studies suggest a full-gap structure \cite{Holenstein16PRB, Kirschner16PRB}, while heat-capacity and heat-transport studies a nodal gap structure \cite{Xing16PRB, Ying16PRB}.
%A STM/STS study supports a highly anisotropic gap structure with a possible node \cite{YangXiong16PRB}.
In both respects, detailed investigations into the electronic structure of FeS are vital.

Our previous de Haas-van Alphen (dHvA) and Shubnikov-de Haas (SdH) measurements on FeS found two low frequencies of quantum oscillations \cite{Terashima16PRB_FeS}.
The angular dependences suggested that they were from (a) Q2D cylinder(s) of the Fermi surface (FS).
Large anisotropy of upper critical field $B_{c2}$ \cite{Borg16PRB, Lin16PRB, Terashima16PRB_FeS} and results of angle-resolved photoemission spectroscopy (ARPES) \cite{Miao17PRB, Reiss17PRB} also support the Q2D electronic structure.
There are however recent dHvA and SdH measurements reporting many other frequencies that cannot be ascribed to Q2D FS cylinders \cite{Man17npjQM}. 
In the present study, we perform dHvA measurements up to $B$ = 45 T, which is more than two times higher than in our previous study \cite{Terashima16PRB_FeS}, and observe eight fundamental frequencies.
We perform quasi-particle self-consistent \textit{GW} (QS\textit{GW}) calculations of the electronic band structure and explain the experimental frequencies quantitatively with the Fermi surface consisting of two hole and two electron cylinders.

\section{Methods}
Tetragonal FeS single crystals were prepared by a hydrothermal method as described in \cite{Lin16PRB}.
The $c$-axis parameters were determined from 00$l$ diffractions to be $c$ = 5.043(9) and 5.042(7) \AA~ for samples B and C, respectively, which are in excellent agreement with previous reports \cite{Lai15JACS, Pachmayr16ChemCommun, Borg16PRB, Lin16PRB}.
(For sample A, only 001 diffraction was observed, which gave $c \sim$ 5.08 \AA.)

DHvA oscillations in the magnetic torque $\tau$ were measured using piezoresistive microcantilevers \cite{Ohmichi02RSI}. 
The 45-T hybrid or a 35-T resistive magnet and a $^3$He refrigerator were used.
The field direction $\theta$ was measured from the $c$ axis.
For a purely two-dimensional FS cylinder, there will be a single dHvA frequency $F$, and $F\cos\theta$ remains constant as the field direction $\theta$ is varied.
If a cylinder is warped by $c$-axis interactions, there will be two frequencies corresponding to the maximum and minimum cross sections, and their $F\cos\theta$ will vary with $\theta$ according to the warping.

In order to interpret dHvA data, we have performed the
first-principles band structure calculations. We have employed two
types of all-electron full-potential methods; the full-potential LAPW
method implemented in the WIEN2K package \cite{WIEN2K} with the
Perdew-Burke-Ernzerhof exchange-correlation functional \cite{Perdew96PRL}, and the
full-potential QS\textit{GW} method \cite{Schilfgaarde06PRL, Kotani14JPSJ} implemented in the ecalj package \footnote{https://github.com/tkotani/ecalj}
with the non-local exchange correlation potential,
\[
V^{xc}_{ij}={1\over2}{\rm
Re}\big[\Sigma_{ij}(\epsilon_i)+\Sigma_{ij}(\epsilon_j)\big],
\]
where $\Sigma_{ij}(\omega)$ is the \textit{GW} self-energy, and $\epsilon_i$ is
the quasiparticle energy at eigenstate $i$.
In both calculations, the experimental lattice parameters were used
\cite{Lai15JACS}, and the spin-orbit coupling was included.
$RK_{max}=7$ and the muffin-tin radii were 2.3 and 1.9 a.u. for Fe and S, respectively, and
$8\times8\times8$ $k$-meshes in the first Brillouin zone were adopted.
Q-mesh grid in the QS\textit{GW} calculations was also $8\times8\times8$.
Along the lines of \cite{Kotani14JPSJ}, cutoff energies for the augmented plane waves and the self-energy were 3.0 Ry. The plane-wave cutoff $|q+G|^\psi_{\rm MAX}=4.0$ and $|q+G|^W_{\rm MAX}=3.0$ in units of bohr$^{-1}$. In the self-consistent calculations, relative error of charge density was smaller than $10^{-5}$.

\section{Results and discussion}
\begin{figure}[!]
\includegraphics[width=8.6cm]{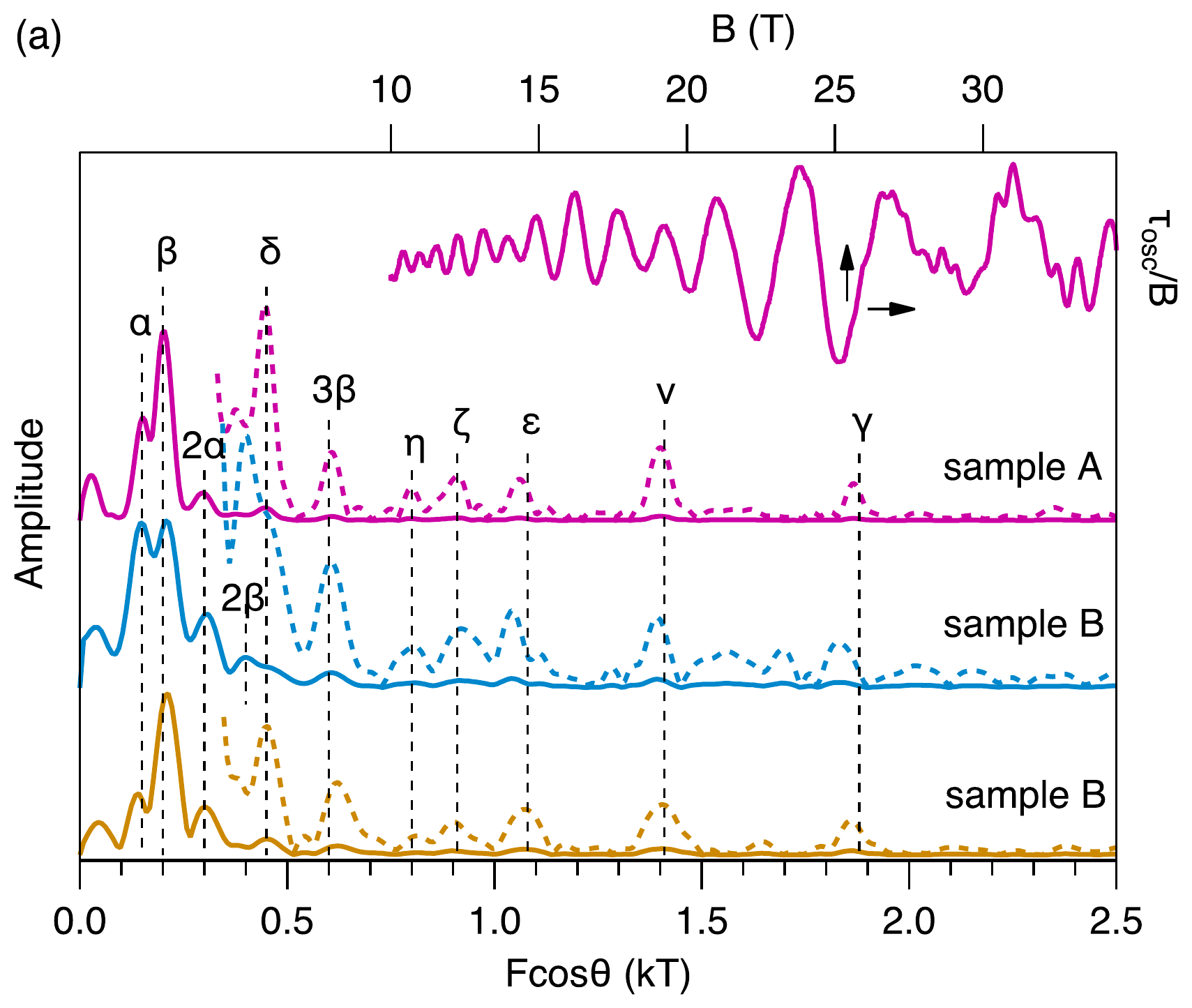}
\includegraphics[width=8.6cm]{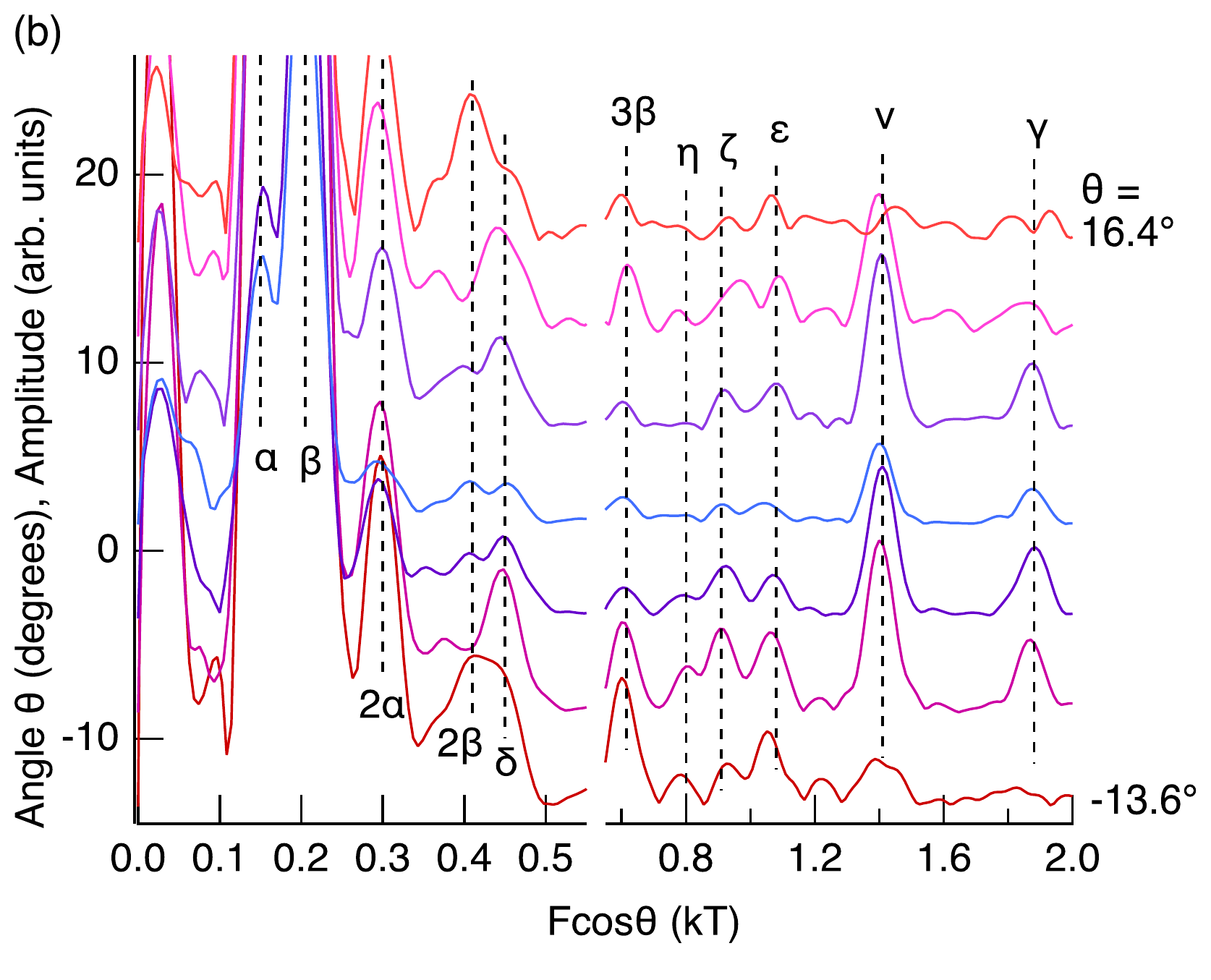}
\caption{\label{sigFT}(Color online).  (a) Fourier transforms of $\tau_{osc}/B$ vs $1/B$ for samples A, B and C.  Note that the horizontal axis is $F\cos\theta$.  $\theta$ = -8.6, -18.3 and -10.6$^{\circ}$ for samples A, B, and C, respectively.  The field range of the transformation is 14 -- 35 T (sample A) or 20 -- 45 T (B and C).  The top right inset shows $\tau_{osc}/B$ vs $B$ for sample A.  (b) Angle dependence of Fourier spectra for sample A.  Note that the horizontal axis is $F\cos\theta$.  The low- and high-frequency parts are based on Fourier transforms in field windows 14 -- 35 T and 20 -- 35 T, respectively.  The spectra are vertically shifted so that the baseline of a spectrum for an angle $\theta$ is placed at $\theta$ of the vertical axis.}   
\end{figure}

\begin{figure}
\includegraphics[width=8.6cm]{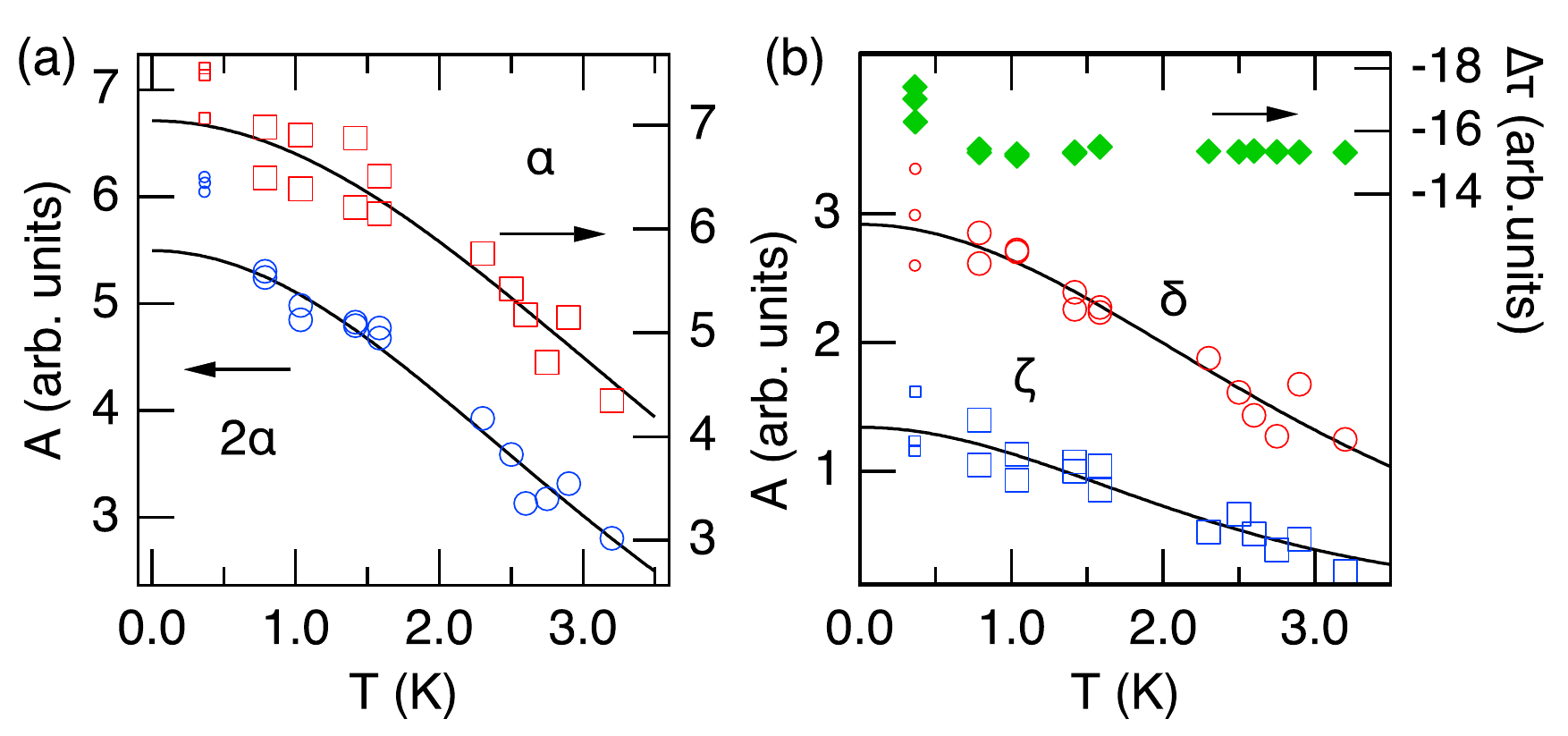}
\caption{\label{mass}(Color online).  Temperature dependence of oscillation amplitudes for the $\alpha$, 2$\alpha$, $\delta$, and $\zeta$ oscillations at $\theta$ = -8.6$^{\circ}$ in sample B.  Field ranges used for the amplitude estimation are 14 -- 45 T for $\alpha$, 20 -- 45 T for 2$\alpha$ and $\delta$, and 25 -- 45 T for $\zeta$.  The solid lines are fits to $R_T= x/\sinh(x)$ where $x=K(m^*/m_e)T/B$ with $K$ = 14.69 T/K in the SI.  The lowest-temperature points indicated by small marks were excluded from the fits.  The diamond marks in the topmost part of (b) show the temperature dependence of $\Delta\tau = \tau(45 \mathrm{~T}) - \tau(11.4 \mathrm{~T})$.}   
\end{figure}

The top right inset of Fig.~\ref{sigFT}(a) shows an example of the oscillatory part $\tau_{osc}$ of the magnetic torque divided by $B$ for sample A.
For a background subtraction, third or fourth order polynomial fitting was used.
The main panel shows Fourier transforms of the torque dHvA oscillations near $B \parallel c$ for samples A, B and C.
They are fairly similar to the Fourier transform of SdH oscillations reported by Man \textit{et al.} (Fig. 4b of \cite{Man17npjQM}) but are distinct from that of torque dHvA oscillations by the same authors (Fig. 4a of \cite{Man17npjQM}).
We identify eight fundamental frequencies $\alpha$, $\beta$, $\delta$, $\eta$, $\zeta$, $\epsilon$, $\nu$, and $\gamma$ (the labels $\delta$, $\epsilon$, $\nu$, and $\gamma$ are after \cite{Man17npjQM}).

It has been reported that magnetic properties of tetragonal FeS are sensitive to details of synthesis conditions \cite{Borg16PRB, Holenstein16PRB, Kirschner16PRB, Kuhn17PhysicaC}.
Indeed, some of our hydrothermally-synthesized crystals were strongly magnetic and did not show superconductivity.
However, more importantly, a total of 13 superconducting samples, including samples A, B, and C, that we measured up to 35 or 45 T \textit{all} showed dHvA oscillations, and dHvA frequencies are in excellent agreement among them.
For example, the $\alpha$ and $\beta$ frequencies were observed for all the samples and the standard deviations of the measured frequencies were less than 1\%.
The standard deviation for the $\nu$ frequency observed for 12 samples was 0.6\%.
Since dHvA frequencies measure sizes of the Fermi surface, they are directly linked with the carrier density and hence the chemical composition.
These observations indicate that superconductivity is an intrinsic property of high-quality tetragonal FeS with a stable composition.

\begin{figure*}
\includegraphics[width=17cm]{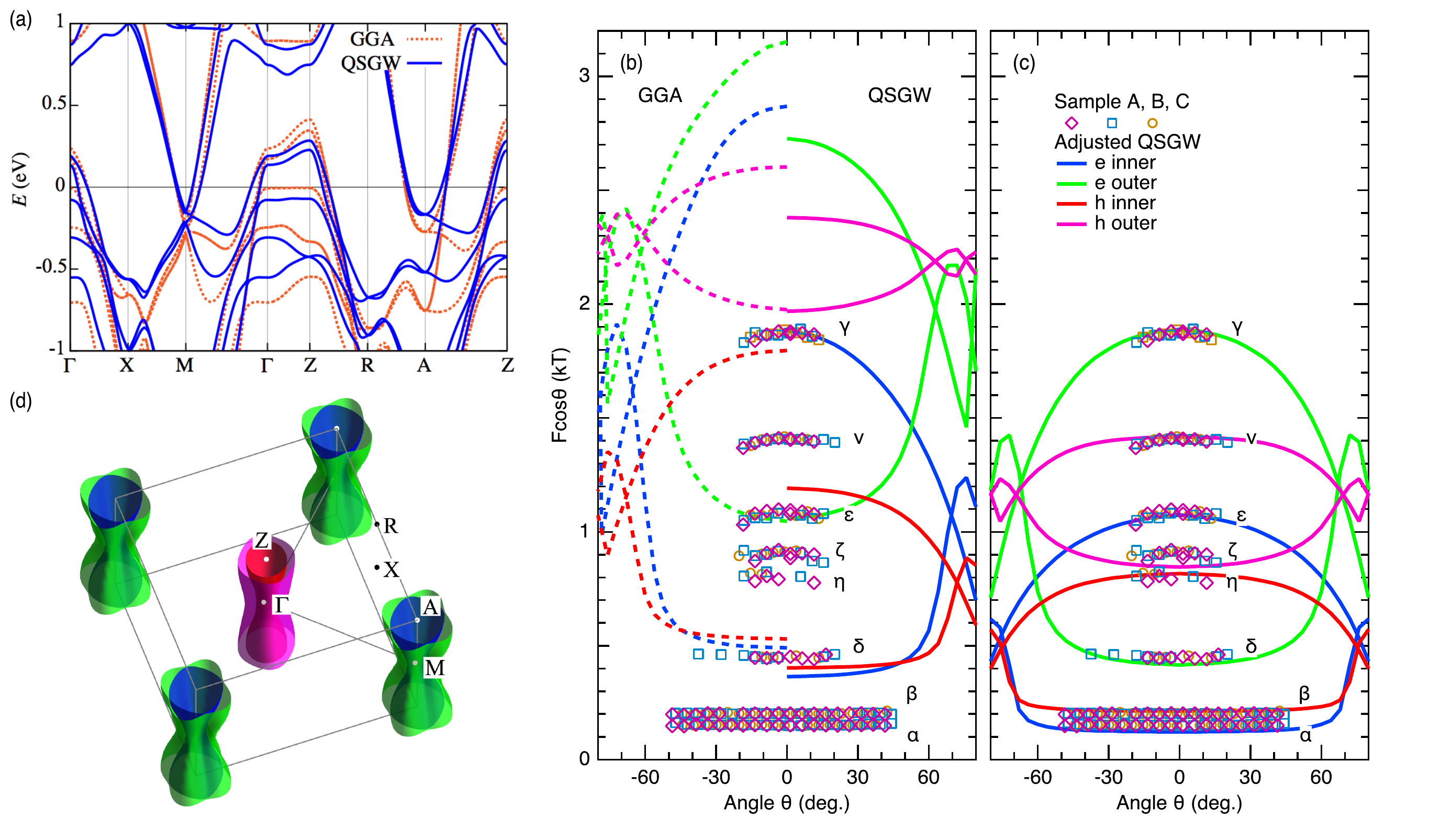}
\caption{\label{FcosAng}(Color online).  (a) Calculated electronic band structures with GGA and QS\textit{GW} approximations.  (b) and (c) Experimental (symbols) vs calculated (lines) dHvA frequencies.  GGA and QS\textit{GW} calculations without band-energy adjustments are shown in (b), while QS\textit{GW} calculations with band-energy adjustments in (c).  Note that the vertical axis is $F\cos\theta$.  (d) Fermi surface resulted from the band-energy adjusted QS\textit{GW} calculations.}   
\end{figure*}

Figure~\ref{sigFT}(b) shows Fourier spectra of dHvA oscillations in sample A for various field directions near $B \parallel c$.
Note that the horizontal axis is $F\cos\theta$.
The peaks corresponding to the eight fundamental frequencies and some of their harmonics are observed for some angle ranges, and $F\cos\theta$ for those peaks stays nearly constant, indicating that they are from Q2D FS cylinders (for $\alpha$ and $\beta$, see also \cite{Terashima16PRB_FeS}).
Consistent results are obtained for samples B and C, and the fundamental frequencies observed in the three samples are plotted in Figs.~\ref{FcosAng}(b) and (c).
Man \textit{et al.} observed many other dHvA frequencies in their dHvA oscillations even for field directions close to $\theta$ = 90$^{\circ}$ \cite{Man17npjQM}, but we could not confirm them.

Figures~\ref{mass}(a) and (b) show the temperature dependence of oscillation amplitudes for the $\alpha$, 2$\alpha$, $\delta$, and $\zeta$ oscillations in sample B.
The solid lines are fits to the Lifshitz-Kosevich formula \cite{Shoenberg84}, which yield effective masses $m^*$ of 0.76(4), 1.25(4), 1.47(7), and 2.2(2) in units of the free electron mass $m_e$, respectively.
The lowest-temperature data points shown by small marks have been excluded from the fits.
The reason is as follows:
The diamond symbols in the topmost part of Fig.~\ref{mass}(b) show the temperature dependence of the magnetic torque.
In order to suppress the influence of a drift in electronics, the difference $\Delta\tau = \tau(45 \mathrm{~T}) - \tau(11.4 \mathrm{~T})$ is plotted instead of $\tau(45 \mathrm{~T})$ ($B$ = 11.4 T is the lowest field for the hybrid-magnet operation).
$\Delta\tau$ exhibits an anomalous increase at the lowest temperature.
Concomitantly, some oscillation amplitudes deviate from extrapolation from high-temperature data.
A similar anomaly has been observed for samples B and C.
On the other hand, we did not see such an anomaly in our previous study \cite{Terashima16PRB_FeS}.
Although the origin of the anomaly is not clear at present, it might be related to trace magnetic impurities in the samples.

\begin{table*}%[H] add [H] placement to break table across pages
\caption{\label{Tab1}Experimental dHVA frequencies and effective masses for $B \parallel c$ compared to the calculated values.  Also shown are orbit areas $A$, Fermi momentums and effective Fermi energies estimated using the following formulas $F=\hbar A/(2\pi e)$, $A=\pi k_F^2$, and $E_F= \hbar^2 k_F^2/(2m^*)$. The experimental values are the averages over the three samples, except that those for $\delta$ are the averages over two samples.  The calculated values are for the adjusted energy bands (see text).  For the orbit assignments, e (h) refers to electron (hole), superscript i (o) inner (outer), and subscript min (max) minimum (maximum) cross section.  The last column shows the mass enhancements.}
\begin{ruledtabular}
\begin{tabular}{cccccccccc}
\multicolumn{6}{c}{Experiment} & \multicolumn{3}{c}{Adjusted calc.} &  \\
\cline{1-6} \cline{7-9}
Branch & $F$ (kT) & $m^*/m_e$ & $A$ (\%BZ) & $k_F$ (\AA$^{-1}$) & $E_F$ (meV) & Orbit & $F$ (kT) & $m_{band}/m_e$ & $m^*/m_{band}$\\
\hline
$\alpha$ & 0.15 & 0.67(5) & 0.50 & 0.068 & 27 & e$^{\mathrm{i}}_{\mathrm{min}}$ & 0.12 & 0.30 & 2.2\\
$\beta$ & 0.20 & 0.85(4) & 0.66 & 0.078 & 27 & h$^{\mathrm{i}}_{\mathrm{min}}$ & 0.21 & 0.35 & 2.4\\
$\delta$ & 0.44 & 1.53(6) & 1.4 & 0.12 & 33 & e$^{\mathrm{o}}_{\mathrm{min}}$ & 0.42 & 0.69 & 2.2\\
$\eta$ & 0.80 &  & 2.6 & 0.16 & &  h$^{\mathrm{i}}_{\mathrm{max}}$ & 0.82 & 0.67 & \\
$\zeta$ & 0.91 & 1.96(8) & 3.0 & 0.17 & 54 & h$^{\mathrm{o}}_{\mathrm{min}}$ & 0.85 & 1.32 & 1.5\\
$\epsilon$ & 1.08 & 2.33(7) & 3.5 & 0.18 & 54 & e$^{\mathrm{i}}_{\mathrm{max}}$ & 1.07 & 1.26 & 1.8\\
$\nu$ & 1.41 & 2.39(3) & 4.6 & 0.21 & 68 & h$^{\mathrm{o}}_{\mathrm{max}}$ & 1.42 & 1.12 & 2.1\\
$\gamma$ & 1.87 & 2.11(6) & 6.1 & 0.24 & 102 & e$^{\mathrm{o}}_{\mathrm{max}}$ & 1.89 & 1.09 & 1.9\\
% Lines of table here ending with \\\textit{\textsf{\textit{}}}
\end{tabular}
\end{ruledtabular}
%\footnotetext[1]{Estimated from the second harmonic data.}
\end{table*}

Table I shows frequencies and effective masses converted to values at $B \parallel c$ assuming $1/\cos\theta$ variation and averaged over the three samples.
Man \textit{et al.} \cite{Man17npjQM} identified our 2$\alpha$ as a fundamental frequency ($\kappa$).
The averaged effective mass for the 2$\alpha$ oscillation was $m^*_{2\alpha}$ = 1.12(7) $m_e$ in the present study, which approximately satisfies the expected relation that $2m^*_{\alpha} = m^*_{2\alpha}$, supporting our assignment.
The fact that $m^*_{2\alpha}$ is slightly smaller than $2m^*_{\alpha}$ can be attributed to the stronger field dependence of the second harmonic than the fundamental:
the effective mass is underestimated when the field dependence of oscillation amplitude is strong.
On the other hand, Man \textit{et al.} identified our $\zeta$ frequency as the second harmonic of $\delta$.
Table I indicates $2F_{\delta} < F_{\zeta}$ and $2m^*_{\delta} > m^*_{\zeta}$.
Those differences are significant and support our assignment.
In addition, the $\eta$ frequency was not reported by Man \textit{et al.}

Before comparing the dHvA data with band-structure calculations, we explain our motivation for QS\textit{GW} calculations.
The QS$GW$ method includes non-local exchange correlation within the \textit{GW} approximation.
It conceptually differs from ordinary self-consistent \textit{GW} in the following sense \cite{Schilfgaarde06PRL, Kotani14JPSJ}:
it is the optimization of the effective one-body Hamiltonian rather than a perturbation treatment starting from the local-density approximation (LDA).
This approach well describes excited-state properties for weakly and moderately correlated materials.
It can accurately predict fundamental gaps in semiconductors, which are known to be underestimated in the LDA or the GGA.
Its applicability is not restricted to semiconductors:
The QS\textit{GW} method has been reported to improve the energy levels of the $e_g$ orbitals in cuprates \cite{Jang15SciRep}.
Further, Tomczak \textit{et al}. have applied the method to iron-based superconductors
and have reported that it can account for some ARPES results better
than LDA/GGA calculations \cite{Tomczak12PRL}.
In the present case of FeS, the band structure is semimetallic, and the electron correlation is moderate.
We can expect that the QS\textit{GW} method produces a better description of the electronic structure than conventional LDA/GGA.

One might argue that the LDA+DMFT (dynamical mean-field theory) method would be more appropriate.
In the iron-based superconductors, it is well-known that the electron/hole pockets in
the LDA/GGA calculations are larger than experimental observations.
It can be partly improved by considering local electron correlation with
the LDA+DMFT method \cite{Anisimov09PhysicaC, Aichhorn09PRB, Yin11NatMat}.
However we note that the DMFT self-energy is local, that is momentum-independent.
On the other hand, the self-energy is momentum-dependent in the QS\textit{GW} method and is expected to directly improve the Fermi-surface shape.
This improvement can be beneficial to analyses of the dHvA data and hence we have chosen the QS\textit{GW} method as the first step.

Figure~\ref{FcosAng}(a) compares the GGA and QS\textit{GW} band structures.
In both cases, the Fermi surface consists of two hole and two electron cylinders at the $\Gamma$ and M points, respectively.
In the GGA band structure, there is a band with $d_{xy}$ character running along the line $\Gamma$Z in the immediate vicinity of the Fermi energy.
Accordingly we suggested in our previous report \cite{Terashima16PRB_FeS} that this band might lie above the Fermi level, producing a small third hole cylinder, and that the $\alpha$ and $\beta$ frequencies might be ascribed to it.
However, the QS\textit{GW} calculation clearly shows that this band sinks below the Fermi level, and hence we abandon this hypothesis.
Comparing the two band structures, we notice the following.
The QS\textit{GW} band structure is more two dimensional than the GGA one: compare the two hole bands along the line $\Gamma Z$ between the two calculations.
The QS\textit{GW} band structure exhibits a smaller band overlap and hence yield smaller FS cylinders.
These features can also be noticed by comparing the dHvA frequencies calculated for the two band structures shown in Fig.~\ref{FcosAng}(b).
The QS\textit{GW} band structure gives smaller frequencies for each FS cylinder, and the difference between the maximum and minimum frequencies at $B \parallel c$ is also smaller.

Figure~\ref{FcosAng}(b) shows that the range of the GGA frequencies already largely overlaps that of the experimental ones.
This is in sharp contrast to the case of FeSe, where the experimental carrier density was more than one order-of-magnitude smaller than predicted by band structure calculations \cite{Terashima14PRB}.
The QS\textit{GW} approximation improves the agreement with the experiment by shrinking the FS cylinders.

We now adjust the QS\textit{GW} band energies to further improve the agreement [Fig.~\ref{FcosAng}(c)].
Band-energy adjustments are often employed in ordinary LDA/GGA analyses of quantum-oscillation data \cite{Terashima11PRL, Coldea08PRL, Carrington11RPP, Analytis09PRL, Arnold11PRB, Putzke12PRL} because there is room for additional band shifts due to the electron correlation effect, mainly, magnetic fluctuations.
The present shifts are similarly justified.
It is reasonable to assume that the $\alpha$ and $\beta$ frequencies correspond to the minimum cross sections of the inner electron [blue in Figs.~\ref{FcosAng}(b) and (c)] and hole (red) cylinders, respectively.
By shifting the inner hole band by -60 meV, the minimum cross section of the inner hole (red) is brought into agreement with $\beta$, and further the maximum one coincides with the $\eta$ frequency [Fig.~\ref{FcosAng}(c)].
For the inner electron (blue), if the band energy is adjusted so that the minimum cross section matches the $\alpha$ frequency, the maximum cross section is brought near to the $\epsilon$ frequency.
We therefore match the maximum cross section to $\epsilon$ with a shift of 74 meV, which brings the minimum cross section into a satisfactory agreement with $\alpha$ [Fig.~\ref{FcosAng}(c)].
It is now reasonable to assume that the two highest frequencies $\gamma$ and $\nu$ correspond to the maximum cross sections of the outer electron (green) and hole (pink) cylinders.
We therefore shift the corresponding band energies by 96 and -90 meV, respectively, to bring them in agreement.
Remarkably, these shifts bring the minimum cross sections into satisfactory agreement with the $\delta$ and $\zeta$ frequencies.
Thus all the experimental frequencies are assigned to FS orbits [Fig.~\ref{FcosAng}(c)] and the determined Fermi surface is shown in Fig.~\ref{FcosAng}(d).
The maximum deviation between the calculated and observed cross sections is less than 0.2\% of the Brillouin zone area for $B \parallel c$.
The most important ingredient of this success is the reduced three dimensionality in the QS\textit{GW} band structure.
In the case of the GGA band structure, because of the stronger three dimensionality, this level of agreement cannot be achieved by momentum-independent band-energy adjustments.

The original QS\textit{GW} band structure gives the carrier densities of 0.0345, 0.0626, 0.0721, and 0.0250 carriers/Fe for the inner and outer electron cylinders and outer and inner hole cylinders, respectively.
After the adjustments, they change to 0.0175, 0.0376, 0.0381, and 0.0157 carriers/Fe.
It is interesting to note that the adjustments improve the overall nesting condition between the holes and electrons:
the carrier densities, which are equivalent to FS volumes, of the inner electrons and holes and also those of the outer electrons and holes become much closer.
The total electron and hole densities are $n_e$ = 0.055 electrons/Fe and $n_h$ = 0.054 holes/Fe, respectively, which satisfy the carrier compensation almost perfectly.
They are about half of the original theoretical values ($n_e$ = $n_h$ = 0.097 carriers/Fe).
The FS volume in iron-based superconductors and their parent compounds with semimetallic character is generally found smaller than predicted by band-structure calculations \cite{Terashima11PRL, Coldea08PRL, Carrington11RPP, Analytis09PRL, Shishido10PRL, Arnold11PRB, Putzke12PRL}.
It is often ascribed to the self-energy due to interband interactions \cite{Ortenzi09PRL}.

The mass enhancement, the ratio of the experimental effective mass to the QS\textit{GW} band mass, does not vary much from orbit to orbit and about two (Table I, last column).
This enhancement is ascribed to the band narrowing due to spin fluctuations that are not sufficiently dealt with in the QS\textit{GW} approximation.
It indicates moderate electron correlations and is comparable to values found in iron phosphides LaFePO, SrFe$_2$P$_2$, BaFe$_2$P$_2$, and LiFeP \cite{Coldea08PRL, Carrington11RPP, Analytis09PRL, Arnold11PRB, Putzke12PRL}.
The density of states after the energy adjustments is calculated to be 0.99 states/eV per spin per cell.
Combining this with the mass enhancement of $\sim$2, the Sommerfeld coefficient is estimated to be $\sim$4.7 mJ/(mol K$^2$), which is in fair agreement with the experimental values of 3.8 or 5.1 mJ/(mol K$^2$) \cite{Xing16PRB, Borg16PRB}.

The nearly perfect carrier compensation and the agreement on the Sommerfeld coefficient substantiate our determined Fermi surface.
This Fermi surface is incompatible with the three-dimensional FS pockets reported in \cite{Man17npjQM}.

The determined Fermi surface has only two hole cylinders and lacks a third cylinder with $d_{xy}$ character.
Within spin-fluctuation paring models, it has theoretically been argued that the existence of the third hole cylinder with $d_{xy}$ character is important in stabilizing a fully-gapped $s_{\pm}$ state \cite{Maier09PRB, Kuroki09PRB, Kemper10NJP, Ikeda10PRB1}.
Thus the preset Fermi surface is favorable for the existence of gap nodes suggested in \cite{Xing16PRB, Ying16PRB, YangXiong16PRB}.

Finally, we comment on ARPES data on FeS.
Miao \textit{et al}. \cite{Miao17PRB} observed two hole and two electron cylinders and their total electron count was a large excess of 0.12 electrons/Fe.
We suggest the excess carriers are likely due to surface effects.
Reiss \textit{et al}. \cite{Reiss17PRB} observed FS cylinders that are much more two-dimensional than ours.
Figure SM3 of \cite{Reiss17PRB} indicates that the Fermi momentum for the inner electron cylinder is 0.10 and 0.12 \AA$^{-1}$ at M and A, respectively.
The corresponding values from the present study are 0.068 and 0.18 \AA$^{-1}$ (Table I).
The discrepancy can be ascribed to the limited $k_z$ resolution of ARPES \cite{Seah79SIA, Strocov02JESRP}.

\section{Summary}
We determined the Fermi surface of tetragonal FeS by combining dHvA measurements and QS\textit{GW} calculations.
The determined FS consists of two hole and two electron Q2D cylinders.
The deviation between the experimental and calculated FS cross-sections is less than 0.2\% of the Brillouin-zone area when momentum-independent band-energy adjustments of less than 100 meV are allowed.
The carrier density is $\sim$0.5 carriers/Fe, and the mass enhancements are $\sim$2.
The nearly-perfect carrier compensation and the fair agreement on the Sommerfeld coefficient substantiate the determined Fermi surface.
The absence of a third hole cylinder is a favorable condition for the formation of nodes in the superconducting gap.
The present Fermi surface of FeS is a starting point to tackle the electronic structure of exotic FeSe.
The wider implications of our results are that the QS\textit{GW} method can be used to derive accurate model Hamiltonians to study paring interactions and symmetries in various iron-based superconductors.
Further, since the DMFT method is complementary to the QS\textit{GW} method with respect to the self-energy,
a combination of QS\textit{GW} and DMFT may provide the comprehensive picture as was previously postulated \cite{Tomczak12PRL}.

\section*{acknowledgments}
TT acknowledges Professors Akira Ochiai and Noriaki Kimura (Tohoku University).
This work was supported by JSPS KAKENHI Grants No. JP17K05556, No. JP17J06088, No. JP16H04021, and No. JP16H01081.
The work in Nanjing University is supported by National Natural Science Foundation of China (NSFC) with the projects: A0402/11534005, A0402/11190023; the Ministry of Science and Technology of China (Grant No. 2016YFA0300404, 2012CB821403).
A portion of this work was performed at the National High Magnetic Field Laboratory, which is supported by National Science Foundation Cooperative Agreement No. DMR-1157490 and the State of Florida.

%\bibliography{/Users/tera/GoogleDrive/Work/bib_files/Pnictides.bib}

%merlin.mbs apsrev4-1.bst 2010-07-25 4.21a (PWD, AO, DPC) hacked
%Control: key (0)
%Control: author (8) initials jnrlst
%Control: editor formatted (1) identically to author
%Control: production of article title (-1) disabled
%Control: page (0) single
%Control: year (1) truncated
%Control: production of eprint (0) enabled
%

\end{document}